\documentclass[twocolumn,aps,floatfix,prl]{revtex4}
\usepackage{amsmath}
\usepackage{graphicx}

\begin{document}
\title{Tunneling-Induced Restoration of the Degeneracy and the Time-Reversal Symmetry Breaking in Optical Lattices}
\author{Tomasz Sowi\'nski$^{1,2,3}$, Mateusz \L\c{a}cki$^{4}$, Omjyoti Dutta$^{2,4}$, Joanna Pietraszewicz$^{1}$, \mbox{Piotr Sierant$^4$},  Mariusz Gajda$^{1,3}$, Jakub Zakrzewski$^{4,5}$, Maciej Lewenstein$^{2,6}$}
\affiliation{
\mbox{$^1$ Institute of Physics of the Polish Academy of Sciences, Al. Lotnik\'ow 32/46, PL-02-668 Warsaw, Poland}
\mbox{$^2$ ICFO --The Institute of Photonic Sciences, Av. Carl Friedrich Gauss, num. 3, E-08860 Castelldefels (Barcelona), Spain  }
\mbox{$^3$ Center for Theoretical Physics of the Polish Academy of Sciences, Al. Lotnik\'ow 32/46, PL-02-668 Warsaw, Poland}
\mbox{$^4$ Instytut Fizyki imienia Mariana Smoluchowskiego, 
Uniwersytet Jagiello\'nski, ul. Reymonta 4, PL-30-059 Krak\'ow, Poland}
\mbox{$^5$ Mark Kac Complex Systems Research Center, 
Uniwersytet Jagiello\'nski, ul. Reymonta 4, PL-30-059 Krak\'ow, Poland   }  
\mbox{$^6$ ICREA -- Instituci{\'o} Catalana de Recerca i Estudis Avan\c{c}ats, Lluis Companys 23, E-08010 Barcelona, Spain} }
\date{\today} 

\begin{abstract}

We study the ground-state properties of bosons loaded into the $p$-band of a
one dimensional optical lattice.  We show that the phase diagram of the system
is substantially affected by the anharmonicity of the lattice potential. In particular,
for a certain range of tunneling strength, the full many-body ground state of the system becomes degenerate. 
In this region, an additional symmetry of the system, namely the parity of the occupation
number of the chosen orbital, is spontaneously broken. The state with nonvanishing staggered angular momentum, which breaks the time-reversal
symmetry, becomes the true ground state of the system.

\end{abstract}
\maketitle

Confining ultracold gases in optical lattices has become a standard experimental tool for studying strongly correlated many-body states
\cite{Jaksch,Greiner}. In the last decade it became possible to arrange experiments in such a~way that they can mimic different ''toy-models'' known
from condensed-matter physics. In this way, optical lattices are truly becoming dedicated quantum simulators for the study of different many-body
configurations \cite{Lewenstein2007,Zwerger}.  In this context, there is a~recent interest in studying ultracold atoms trapped in the higher bands of optical
lattices \cite{Alon2005,Scarola2005,Lewenstein2011}. Such states can be prepared by loading ultracold atoms into the higher bands \cite{Bloch1, Hemmerich1, Hemmerich2,Porto} or via interaction-induced
interorbital transfer \cite{Sengstock, Sowinski, Lacki, Dutta, Pietraszewicz2013}. Atoms in such excited bands can give rise to a~zoo of exotic phases, such as
exotic superfluidity with broken time-reversal symmetry (TRS) \cite{Hemmerich1, Hemmerich2, Porto, Sengstock, Isacsson, Liu2006, Wu2006, Larson2009, Collin2010,Pinheiro2012},
dynamical topological insulators \cite{Dutta}, flat-band crystallization \cite{Wu2007} etc. 

In a theoretical treatment of Bose-Hubbard models the harmonic approximation (HA) is often used to get the first most important insight into the physics of the system. Simplifying the Hamiltonian using a Gaussian basis is the simplest route to checking for the possible phases of the system. One expects that these states remain stable while the single-site symmetries are lowered by the presence of the lattice, i.e. when going beyond the HA approximation. However, it was shown that anharmonicity may have important consequences when one studies $p$-band physics \cite{Collin2010,Pinheiro2012,Pietraszewicz2013}. In our Letter, we demonstrate that the apparently technical and quantitatively small change of harmonically approximated on-site functions to the true orthonormal basis of Wannier states may substantially change the phase diagram and many-body physics in a nonperturbative way. We believe that this finding is general for $p$-band physics and that the results obtained in the HA should be interpreted with care.

As an example, we focus on a~specific model for $p$-orbital bosons in an effective one-dimensional (1D) optical lattice as introduced in Ref. \cite{Liu}. In the HA, when $p_x$ and $p_y$ orbitals are degenerate the system shows anti-ferro-orbital ordering which break TRS in the limit of small tunneling.  This is due to the fact that the local on-site Hamiltonian within the HA commutes with the angular momentum operator for the lattice sites. This is no longer true when one takes into account the full complexity of the problem \cite{1dremark}.

The Hamiltonian describing spinless bosons confined in the optical lattice potential ${\cal V}(\boldsymbol{r})$ interacting via contact interactions is
\begin{align}
\hat{\cal H} = \int\!\!\mathrm{d}\boldsymbol{r}\, \hat\Psi^\dagger(\boldsymbol{r}) \left[-\frac{\hbar^2}{2m}\nabla^2 + {\cal V}(\boldsymbol{r}) + \frac{g}{2}\hat\Psi^\dagger(\boldsymbol{r})\hat\Psi(\boldsymbol{r})\right] \hat\Psi(\boldsymbol{r}).
\end{align}
The field operator $\hat\Psi(\boldsymbol{r})$ annihilates a~boson at point $\boldsymbol{r}$. Parameter $g$ describes the strength of the contact interactions, and it is proportional to the $s$-wave scattering length. We expand the field operator $\hat\Psi(\boldsymbol{r})$ in a single-particle basis of maximally localized Wannier states $\phi_{\boldsymbol{i}}^{\boldsymbol{\alpha}}(\boldsymbol{r})$:
\begin{equation} \label{PSI}
\hat\Psi(\boldsymbol{r}) = \sum_{\boldsymbol{i}}\sum_{\boldsymbol{\alpha}} \hat{a}_{\boldsymbol{\alpha}}(\boldsymbol{i}) \phi_{\boldsymbol{i}}^{\boldsymbol{\alpha}}(\boldsymbol{r}).
\end{equation}
The operator $\hat{a}_{\boldsymbol{\alpha}}(\boldsymbol{i})$ annihilates a~single boson at site $\boldsymbol{i}$ occupying Bloch band $\boldsymbol{\alpha}$ of the periodic potential. In the case of a~rectangular two-dimensional optical lattice, i.e., when 
\begin{equation}
{\cal V}(\boldsymbol{r}) = V_{x}\sin^2(2\pi x/a_x) + V_y\sin^2(2\pi y/a_y) + \frac{m\Omega^2}{2} z^2
\label{poten}
\end{equation}
the single-particle Hamiltonian is separable and Wannier states are products of one-dimensional Wannier functions $\phi_{\boldsymbol{i}}^{\boldsymbol{\alpha}}(\boldsymbol{r})={\cal W}_{i_x}^{\alpha_x}(x){\cal W}_{i_y}^{\alpha_y}(y){\cal Z}_0(z)$, where $\boldsymbol{i}=(i_x,i_y)$ and $\boldsymbol{\alpha}=(\alpha_x,\alpha_y)$. We assume that the dynamics in the $z$ direction is completely frozen and that all bosons occupy the lowest state of the harmonic confining potential ${\cal Z}_0(z)$.  

Note that our potential does not take into account the slowly spatially varying harmonic trap typically present in experiments, which may affect the properties of $p$-band states \cite{Pinheiro2012}. In the following, we shall assume a uniform filling of sites, thus explicitly excluding the additional trapping. With care, such a situation may be realized experimentally \cite{Will2010,Gaunt13}.

The standard Bose-Hubbard Hamiltonian describing the lowest band dynamics is obtained by restricting the decomposition \eqref{PSI} to the lowest band only, i.e. $\alpha_x=\alpha_y = 0$ \cite{Lewenstein2007,Zwerger}. 
Here, we are interested in many-body properties of bosons occupying higher orbitals of the optical lattice. Therefore, we assume that the system is prepared in such a~way that all particles occupy only the first excited Bloch band ($p$ band) of the optical lattice. 
We consider a
highly nonsymmetric lattice with $V_y \gg V_x$. For large enough $V_y$, the tunneling in the $y$ direction is suppressed and an effectively one-dimensional chain is obtained. The 
ratio $a_x/a_y$ between lattice constants is adjusted to preserve single-particle degeneracy between the $p_x$ and $p_y$ orbitals at each lattice site. 
The bosons can  tunnel to neighboring sites along the $x$ direction: the $p_x$ ($p_y$) orbital tunnels with amplitude $t_x<0$ ($t_y>0$) and $|t_x|>|t_y|$. The difference in sign and magnitude of the tunneling amplitudes is a~direct consequence of the fact that tunneling of a particle in $p_y$ ($p_x$) orbital is equal to the tunneling in the ground (excited) Bloch band of the optical lattice in $x$ direction.  
In the HA, the single-site part of the Hamiltonian becomes rotationally invariant since in that case the contact interactions between bosons preserve the local rotational symmetry. In consequence, for small tunneling, the ground state of the system is doubly degenerate and in the thermodynamic limit the system undergoes spontaneous symmetry breaking. 

Here we reexamine the properties of the~system taking into account the anharmonicity (and the resulting anisotropy) of sites. We assume that bosons can occupy only the $p$-band states of the one-dimensional optical chain and we restrict the decomposition \eqref{PSI} to $\boldsymbol{\alpha}=(1,0)$ and $\boldsymbol{\alpha}=(0,1)$. This neglects collisional couplings with other modes, notably the resonant collisions in which bosons from $(0,1)$ and $(1,0)$ modes are transferred to the $(0,0)$ and $(1,1)$ modes \cite{Pietraszewicz2013}.
Hopefully, such collisions can be suppressed in optical lattice experiments similarly to earlier
works \cite{Bloch1,Hemmerich1,Hemmerich2}. Here, we restrict ourselves to $p$-band physics only. The annihilation operators for the two modes considered are denoted as $\hat{a}_x(i)$, $\hat{a}_y(i)$. These operators annihilate bosons in the single-particle Wannier states $\phi_{i}^x(\boldsymbol{r})={\cal W}_{i}^{(1)}(x){\cal W}_{i}^{(0)}(y){\cal Z}_0(z)$ and $\phi_{i}^y(\boldsymbol{r})={\cal W}_{i}^{(0)}(x){\cal W}_{i}^{(1)}(y){\cal Z}_0(z)$, respectively. We also introduce density operators 
$\hat{n}_\alpha(j)=\hat{a}_\alpha^\dagger(j)\hat{a}_\alpha(j)$ ($\alpha=x,y$). 
Then the  Hubbard-like Hamiltonian describing dynamics in the 1D chain is expressed as
\begin{subequations} \label{Ham} 
\begin{align} 
\hat{\cal H} &= \sum_{j} \hat{H}(j)-\sum_{\langle ij\rangle} \left[ t_x \hat{a}_x^\dagger(i) \hat{a}_x(j) + t_y \hat{a}_y^\dagger(i) \hat{a}_y(j)\right].
\end{align}
The local, on-site Hamiltonian $\hat{H}(j)$ has the form
\begin{align} \label{HamLoc}
&\hat{H}(j) =\sum_{\alpha=x,y} \left[E_\alpha \hat{n}_\alpha(j) + \frac{U_{\alpha\alpha}}{2} \hat{n}_\alpha(j)(\hat{n}_\alpha(j)-1)\right] \\
&+\frac{U_{xy}}{2} \left[4 \hat{n}_x(j)\hat{n}_y(j)+\hat{a}_x^\dagger(j)^2\hat{a}_y(j)^2+\hat{a}_y^\dagger(j)^2\hat{a}_x(j)^2\right]. \nonumber
\end{align} 
\end{subequations} 
All $U$'s denote contact interactions between appropriate orbitals. $E_x$ and $E_y$ are single-particle energies, which in general differ. It is obvious that the Hamiltonian commutes with the total particle number operator $\hat{N}=\hat{N}_x+\hat{N}_y$, where $\hat{N}_\alpha = \sum_i \hat{n}_\alpha(i)$. This is a property not enjoyed by  $\hat{N}_x$ and $\hat{N}_y$ separately,  due to the last two terms in the local Hamiltonian that  transfer  {\em pairs} of bosons between different orbitals. Thus the Hamiltonian has global $Z_2$ symmetry related to the parity of the operator $\hat{N}_y$ (choosing $\hat{N}_x$ leads to the same conclusions) and it commutes with the symmetry operator ${\cal S}=\exp(i\pi\hat{N}_y)$. To find the ground state of the system (in the subspace spanned by the $p$-band states) one can find the lowest energy states in the two eigen-subspaces of ${\cal S}$ independently. Let us call these states $|\mathtt{G}_{even}\rangle$ 
and $|\mathtt{G}_{odd}\rangle$ with corresponding eigenenergies $E_{even}$ and $E_{odd}$ [subscripts even (odd) correspond to even (odd) numbers of bosons in orbital $y$]. Finally, one can choose the state with lower energy as the global ground state (GS) of the system. In principle it may happen that both ground states have the same energy. In such a~case, any superposition $\cos(\theta)|\mathtt{G}_{even}\rangle+\sin(\theta)\mathrm{e}^{i\varphi}|\mathtt{G}_{odd}\rangle$ is a~ground state of the system. As explained later, in the thermodynamic limit this $U(1)\times U(1)$ symmetry is spontaneously broken to Ising-like $Z_2$ symmetry, and only one of two macroscopic states can be realized. 

\begin{figure}
\includegraphics{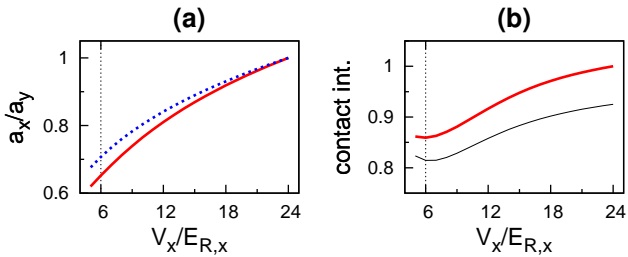}
\caption{(color on-line) Influence of the anharmonicity on the parameters of the Hamiltonian. (a) The ratio $a_x/a_y$ in the HA (dotted blue line) and in the Wannier basis (solid red line) that equalizes single-particle energies $E_x=E_y$. (b) The ratio $U_{xx}/U_{yy}$ (thick red line) and $3U_{xy}/U_{yy}$ (thin black line) calculated for exact Wannier functions. In both plots we take $V_y/E_{R,y}=24$. The dotted vertical line indicates the case $V_x/E_{R,x}=6$ studied in the text. In the HA, both quantities are equal to 1. \label{Fig1}}
\end{figure}

Now, let us discuss the role of the anharmonicity of the lattice potential. To be specific, we take lattice depths $V_x/E_{R,x}=6$, $V_y/E_{R,y}=24$ and filling $\nu=3/2$, where $E_{R,\alpha}=2\pi^2 \hbar^2/ma_\alpha^2$. The single-particle energies $E_x$ and $E_y$, can be equalized , even for different lattice depths by changing the lattice constants $a_x$ and $a_y$. For chosen $V_x, V_y$ one can show in the HA that  $a_x/a_y=1/\sqrt{2}$ leads to equal single-particle energies. Nevertheless, the ratio $a_x/a_y$ calculated directly in the basis of Wannier functions differs from that value (see Fig. \ref{Fig1}(a)), and for the example studied it is approximately equal to $0.65$. 

Importantly, the anharmonicity also dramatically changes the contact interactions. Since the wave functions of $p$ orbitals are products of one-dimensional functions, the ratios $U_{xx}/U_{yy}$ and $U_{xy}/U_{yy}$ do not depend on the lattice constants $a_x$ and $a_y$. They are functions of dimensionless lattice depths $V_x/E_{R,x}$ and $V_y/E_{R,y}$ only.  Moreover, in the HA they are equal to $1$ and $1/3$, respectively, independently of the lattice depths. This simple observation indicates that the HA may be valid only in the very deep lattice regime. It is straightforward to show that in the HA the Hamiltonian Eq. \eqref{Ham} reduces to a Hamiltonian that preserves angular momentum in each lattice site independently. The precise values of the ratios $U_{xx}/U_{yy}$ and $U_{xy}/U_{yy}$ calculated from appropriate Wannier wave functions are presented in Fig. 1(b). We see that whenever dimensionless lattice depths are different, one finds $U_{xx}\neq 
U_{yy}$. Moreover, even for equal lattice depths, the contact interaction $U_{xy}$ is never equal to $U_{xx}/3$. It means that the harmonic limit can not be reached in any realistic optical lattice, and rotational invariance of local lattice sites does not hold. 

The violation of local rotational invariance is clearly visible when we change the single-particle basis from the Cartesian to the angular one. By introducing angular-momentum-like annihilation operators $\hat{a}_\pm(j) = \left[\hat{a}_x(j) \pm i \hat{a}_y(j)\right]/\sqrt{2}$ the local part of the Hamiltonian Eq. \eqref{Ham} can be written in the form
\begin{align}
\hat{H}(j) &= \frac{U}{2} \left[ \hat{n}(j)\left(\hat{n}(j)-\frac{2}{3}\right) - \frac{1}{3}\hat{L}_z^2(j)\right]  \nonumber \\
&+ \delta\left[\left(\hat{n}(j)-1\right)\left(\hat{L}_+(j)+\hat{L}_-(j)\right)\right] \nonumber \\
&+\lambda\left[\frac{1}{4}\hat{L}_z^2(j) -3\left(\hat{L}_+(j)-\hat{L}_-(j)\right)^2-\hat{n}(j)\right]
\end{align}
where $U = (U_{xx}+U_{yy})/2$, $\delta=(U_{xx}-U_{yy})/2$, and $\lambda=U_{xy}-U/3$ with   $\hat{n}(j) = \hat{a}_+^\dagger(j)\hat{a}_+(j)+\hat{a}_-^\dagger(j)\hat{a}_-(j)$, and angular momentum operators $\hat{L}_z(j)=\hat{a}^\dagger_+(j)\hat{a}_+(j) - \hat{a}^\dagger_-(j)\hat{a}_-(j)$, $\hat{L}_\pm(j) = \hat{a}^\dagger_\pm(j)\hat{a}_\mp(j) /2$. 
In the HA  $\lambda=\delta=0$ for any lattice parameters
$[\hat{H}(j), \hat{L}_z(j)]=0$, and the 
eigenvalues of $\hat{L}_z(j)$ become good quantum numbers. However, for the optical lattice potential \eqref{poten}, we find that $\lambda, \delta \neq 0$, and consequently $[\hat{H}(j), \hat{L}_z(j)] \neq 0$,  breaking the local axial symmetry.

\begin{figure}
\includegraphics{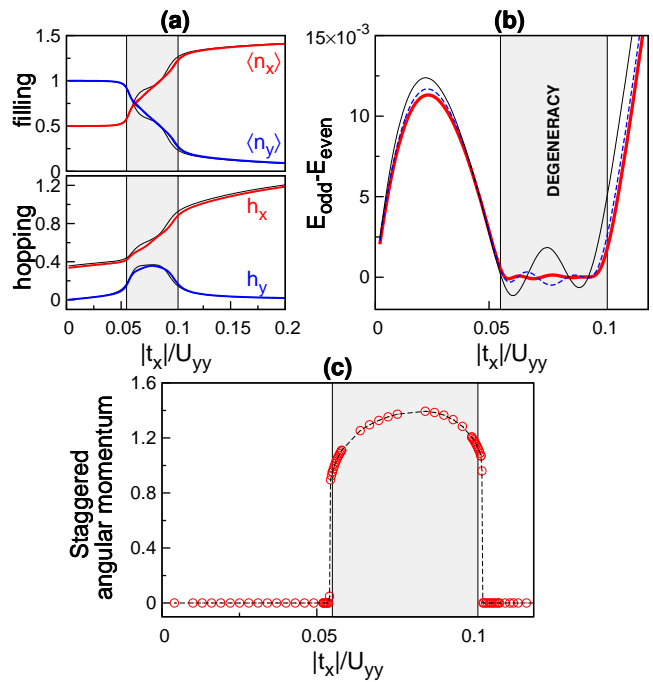}
\caption{(color on-line) (a) Filling and hopping of the $p_x$ (red line) and $p_y$ (blue line) orbitals for $\nu = 3/2$ obtained with the ED method on a~lattice with $L=6$ sites. Results agree with corresponding results obtained for $L=4$ (thick black lines) and DMRG calculations (not shown since practically indistinguishable from ED data). (b) The energy difference between the two ground states in even and odd subspaces of the eigenstates of the symmetry operator ${\cal S}$. The energies are obtained with the ED method on the lattice with $L=4,6$, and $8$ sites (thin black, dashed blue, and thick red lines, respectively). Note that corresponding lines cross the zero energy  $L$ times. (c) Expectation value of the staggered angular momentum  $\hat{\cal L}_z/L$ as a~function of tunneling obtained with DMRG on the lattice with $L=64$ sites. Nonvanishing value of $\hat{\cal L}_z$ is present only in the region where the ground state is degenerate. In all figures the shaded region denotes the range of tunnelings where the ground state of the system 
is degenerate in the thermodynamic limit.\label{Fig2}}
\end{figure}

The properties of the ground state crucially depend on the contact interactions. As an example, we revisit the case of total filling $\nu = 3/2$. We performed exact diagonalization (ED) in the full many-body basis (the basis formed by all Fock states $|n_1,\ldots,n_L\rangle$ with $\sum_i n_i = \nu L$) on the lattice with $L=4,6$, and $8$ sites and periodic boundary conditions. In the HA we find that for small tunnelings, the GS of the system is degenerate, i.e. both ground states $|\mathtt{G}_{even}\rangle$ and $|\mathtt{G}_{odd}\rangle$ have the same energy. In this way we reproduce the results of the HA. However, when the anharmonicity is taken into account, the GS looses its degeneracy [Fig. \ref{Fig2}(a)]: in the limit of small tunnelings, the GS becomes the insulating state in the $p_y$ orbital with one boson per site and the fractional superfluid state in the $p_x$ orbital. Moreover, we do not find any significant correlations $\langle a_x^\dagger(j)a_y(j)\rangle$ in this limit. In contrast, we find that for large tunneling all particles occupy the $p_x$ orbital in the superfluid phase. This is manifested by a large hopping correlation $h_x$
\begin{equation} \label{hopping-def}
h_\alpha = \frac{1}{L}\sum_j\langle a_\alpha^\dagger(j)a_\alpha(j+1)\rangle,
\end{equation}
where $\alpha = x, y$. These results were confirmed for a larger $L=64$ system using a density matrix renormalization group (DMRG) approach \cite{White,italians}.

The most interesting scenario is realized for intermediate values of the tunneling. As is visible in Fig. \ref{Fig2}(a), the balance $N_x-N_y$ has the opposite sign in the two limiting cases ($t_x\rightarrow 0$ and $|t_x|\rightarrow\infty$). Therefore, there exists a particular tunneling value for which both orbitals are balanced. To check this point we plot the energy difference between ground states $E_{odd}-E_{even}$ as a~function of tunneling for different lattice sizes [Fig. \ref{Fig2}(b)]. We find that near the balanced tunneling point both ground states are degenerate. In fact, increasing the lattice size $L$, we find that degeneracy occurs for exactly $L$ different values of the tunneling within a certain finite range. The range of tunneling for which $E_{odd}-E_{even}=0$ does not grow with lattice size, but saturates.  Because of this observation, we claim that in the thermodynamic limit the degeneracy of the ground state is recovered in a certain well-defined range of tunnelings. In this region, whenever the tunneling is changed, one particle is transferred between orbitals to minimize the energy. Since there is no corresponding term in the Hamiltonian, this transfer is directly related to the flip from one eigenspace of ${\cal S}$ to the other. 

In the region of tunneling-induced degeneracy both ground states $|\mathtt{G}_{even}\rangle$ and $|\mathtt{G}_{odd}\rangle$ have the same energy. However, in the thermodynamic limit, due to the einselection principle \cite{Zurek}, the macroscopic state that is realized physically should exhibit as low an entanglement as possible. To find this state, we look for such a~combination of ground states in which the entanglement entropy for one lattice site is the lowest. We minimize the von Neumann entropy defined as $S(\theta,\varphi) = -\sum_i \lambda_i \ln \lambda_i$ as a~function of angles $\theta$ and $\varphi$, where the $\lambda$ values are the eigenvalues of the reduced density matrix for a single lattice site.  With this procedure we find two orthogonal ground states $|\mathtt{G}_\pm\rangle = (|\mathtt{G}_{even}\rangle\pm i|\mathtt{G}_{odd}\rangle)/\sqrt{2}$ with the lowest entropy. For these states 
the reduced density matrix has two dominant eigenvalues $\lambda_1=\lambda_2 \sim 1/2$, which vary only insignificantly with the tunneling. The same two states minimize the von Neumann entropy of a subsystem of two lattice sites. There, the reduced density matrix has three dominant eigenvalues $\lambda_1=\lambda_2\sim 1/3$ and $\lambda_3\sim 1/6$.

This situation is very similar to the situation in the standard Ising system in transverse field. In that case, the system also has the additional symmetry of flipping all spins. After diagonalization of the Hamiltonian in two independent eigen-subspaces of the symmetry operator, one finds two degenerate ground states $|\pm\rangle = (|\mathtt{UP}\rangle\pm|\mathtt{DOWN}\rangle)/\sqrt{2}$, where $|\mathtt{UP}\rangle$ and $|\mathtt{DOWN}\rangle$ denote states with all spins up and down, respectively. 
In the thermodynamic limit, due to the einselection principle, 
the symmetry of the ground state is spontaneously broken and only the $|\mathtt{UP}\rangle$ or $|\mathtt{DOWN}\rangle$ state can be physically realized. From this perspective, our system also has Ising-like $Z_2$ symmetry connected to the symmetry operator ${\cal S}$. This symmetry is spontaneously broken in the thermodynamic limit to one of the 
two states $|\mathtt{G}_\pm\rangle$. Note, however, that in contrast to the usual Ising model, 
the broken symmetry  states are {\em complex} superpositions.
This is a very unusual situation since the original Hamiltonian in the Cartesian basis as well as in the angular momentum basis is represented by a purely real matrix. Complex superpositions appear only due to the additional assumption that ''Schr\"odinger cat'' states can not be obtained in the thermodynamic limit. This assumption has nontrivial consequences. The macroscopic ground state has nonvanishing correlations $C_{\alpha\beta}(j)=\langle a_\alpha^\dagger(j)a_\beta(j)\rangle$ for $\alpha\neq\beta$ 
(being a superposition of states with different $N_x$ values). 
In addition, the nontrivial complex factor in the ''superposed'' state is responsible for a sign inversion symmetry of the correlations  $C_{xy}(j)=-C_{yx}(j)$. These facts lead directly to the observation that the broken symmetry ground state violates TRS since this is the state with a nonvanishing expectation value of staggered angular momentum operator $\hat{\cal L}_z = \sum_j (-1)^j \hat{L}_z(j)$ [Fig. \ref{Fig2}(c)]. 

To conclude, we have shown that anharmonicity of the optical lattice sites plays a~crucial role when one studies the orbital properties of the system. The assumption of harmonicity for lattice sites is highly oversimplified,  since it leads directly to a rotational invariance of the local Hamiltonian in all possible arrangements of the optical lattice. We show that in  asymmetric rectangular lattices it is possible to obtain degeneracy between the single-particle energies of $p$ orbitals by a proper adjustment of the lattice parameters. However, this degeneracy of the single-particle levels is always lifted by anharmonicity when contact interactions are taken into account. Instead, we identify another symmetry operator corresponding to eigenstates containing even and odd numbers of particles in specific $p$ orbitals. We find that the degeneracy between them is dynamically restored due to tunneling. Additionally, we show that in the thermodynamic limit, the ground state breaks TRS along with the $U(1)\times U(1)$ symmetry down to Ising-like $Z_2$ symmetry. Examples of such tunneling-induced restoration of the degeneracy and resultant breaking of symmetry are quite rare in condensed-matter physics, though such cases are known in high-energy physics (i.e., the Schwinger mechanism for dynamical generation of mass). 
The studied system can be probed experimentally by trapping ultracold atoms in an optical lattice setup.

We are grateful to P. Deuar for his fruitful comments and suggestions and G.~de~Chiara, D.~Rossini, and S.~Montangero for help with the DMRG code released within the ``Powder with Power'' project (qti.sns.it). DMRG simulations were performed using the PL-Grid Infrastructure via ACK Cyfronet AGH. The work was supported by (Polish) National Science Center Grants No. DEC-2011/01/D/ST2/02019 (TS, JP), DEC-2011/01/N/ST2/02549 (M\L), DEC-2011/01/B/ST2/05125 (MG), and DEC-2012/04/A/ST2/00088 (OD, JZ). The Spanish MINCIN project TOQATA (FIS2008-00784), ERC Advanced Grant QUAGATUA, EU IP SIQS, and EU IP AQUTE are also acknowledged for generous support. T.S. acknowledges support from the Foundation for Polish Science (KOLUMB Programme; KOL/7/2012).

\end{document}